\documentclass[12pt,a4paper]{article}
\usepackage{amssymb,amsfonts,amsmath}
\usepackage{color}
\usepackage{epsfig, euscript, graphics}

\begin{document}

\title{Social laser model for the Bandwagon effect: generation of coherent information waves}

\author{Andrei Khrennikov\\  Linnaeus University, International Center for Mathematical Modeling\\  in Physics and Cognitive Sciences
 V\"axj\"o, SE-351 95, Sweden}

\maketitle

\abstract{During the last years our society was often exposed to the coherent information waves of high 
amplitudes. These are waves of huge social energy. Often they are of the destructive character, a kind of information tsunami. But, they can carry as well positive improvements in the human society, as waves of decision making matching rational recommendations 
of societal institutes.  The main distinguishing features of these waves are their high amplitude, coherence (homogeneous character of social actions generated by them), and  short time needed for their generation and relaxation. 
Such waves can be treated as large scale exhibition of the Bandwagon effect.
We show that this socio-psychic phenomenon can be modeled on the basis of the recently developed {\it social laser theory}.  This theory can be 
used to model {\it stimulated amplification of coherent social actions}. ``Actions'' are treated very generally, from mass protests to votes and other collective decisions, as, e.g., acceptance (often unconscious) of some societal recommendations. In this paper, we concentrate on theory of laser resonators, physical vs. social. For the latter, we analyze in very detail functioning of the internet based Echo-Chambers. Their main purpose is increasing of the power of the quantum information field as well as its coherence. Of course, the Bandwagon effect is well known and well studied in social psychology. However, the social laser theory gives the 
possibility to model it by using the general formalism of quantum field theory. The paper contains minimum of mathematics and it can be readable by researchers working in psychology, cognitive, social, and political sciences; it might also be interesting for experts in information theory and artificial intelligence.} 

{\bf keywords:} coherent information waves, social energy, quantum-like modeling in decision making, social laser model, resonator, Echo Chamber.

\section{Introduction}

During the last years the grounds of the modern world were shacked by the coherent information waves of very high amplitude. The basic 
distinguishing property of such waves is that they carry huge amount of {\it social energy.} Thus, they are not just the waves widely distributing  some special information content throughout the human society. In contrast, their information contents are very restricted. Typically, the content carried by a wave  is reduced to one (or a few) labels; we can say ``colors'': one wave is ``green'', another is ``yellow''. At the same time information waves carry very big emotional charge, a lot of social energy. So, they can have strong destructive as well as constructive impact on the human society.  In this note, we present a model of generation of very powerful and coherent information waves; the model based on the recently developed theory of {\it social laser} \cite{L1}-\cite{LX}. 

We stress that social laser is a part of the extended project on applications of the formalism of quantum theory outside of physics,
{\it quantum-like modeling} (see, e.g., monographs \cite{QL0}-\cite{QL4} and some selection of papers \cite{PI1}-\cite{z3}). This terminology was invented by the author to distinguish this modeling from attempts to reduce human 
consciousness, cognition, and consequently behavior to genuine quantum physical processes in the brain (see, e.g., Penrose \cite{P} 
or Hameroff \cite{H}). We do not criticize the genuine quantum physical approach to cognition (and consciousness). Some insights 
from  it were useful for the social laser project; in particular, the latter was influenced  by
works on the quantum field theory of cognition, by Ricciardi, Umezawa \cite{RU} and Vittiello \cite{V0,V1}.

Previously, the social laser theory was used to model {\it Stimulated Amplification of Social Actions} (SASA) such as color revolutions and other mass protests around the world (cf. with sociopolitical 
studies, e.g.  \cite{CR1}-\cite{CR8}). However, it is clear that the social laser theory has essentially wider domain of applications, as we shall see in this work. Another aim of this work is present the nutshell of the social lasing theory with 
minimal appealing to the mathematical formalism (cf. with the formal presentation in previous works \cite{L1}-\cite{LX}). We hope that 
this presentation would be useful for researchers with interests in human psychology, decision making, social, cognitive, and information sciences who do not have any background in quantum formalism. 

 It is  also useful to remark that Haken (one of the creators of the laser theory) considered  laser-equations to illustrate the mathematical analogy with self-organization processes in complex physical, biological, and social systems \cite{H0}-\cite{HX} (see also 
\cite{LS}). Our aim is different. We want to formalize quantum features of  information systems (including humans) which can lead to generation of big information waves. The waves having very high degree of coherence, i.e., homogeneity with respect to communications' content.

Nowadays, the mathematical formalism and the methodology of quantum theory are widely used in psychology, decision making, cognitive, social, and political sciences, game theory, economics and finance\footnote{See, e.g., \cite{QL0}-\cite{z3} and references herein and in aforementioned  monographs. Coupling of this paper to quantum-like  decision making is not straightforward. Therefore we do not even try to present more or less complete review on this topic.}. There is a plenty of statistical data confirming that the quantum-like probabilistic model match these data better than the classical one. These data was originally collected without any relation to quantum-like modeling, mainly in cognitive psychology, behavioral economics and finance, and  game theory. Such statistical data was connected to irrational behavior of humans, see, e.g., the pioneer works of 
Kahneman (the Nobel prize in behavioral economics) and Tversky \cite{Ka1,Ka2}. Later, it became clear that these irrationality related data can be consistently modeled with the aid of the probability counterpart of the  quantum formalism \cite{PI3,IR2}. In particular, this approach resolved 
all basic paradoxes of classical decision theory such as Allais (1953), Ellsberg (1961) or Machina (2009) paradoxes \cite{PA,PE,PM}.

The social laser theory formalized the basic conditions for successful lasing \cite{L4}:
\begin{itemize}
\item {\it Indistinguishability of people.} The human gain medium, population exposed to the information radiation, should be composed of so to say {\it social atoms}, ``creatures without tribe'': the role of national, cultural, religious, and even gender differences should be reduced as much as possible. 
\item {\it Content-ignorance.} Social atoms should process information communications without deep analyzing of their contents; 
they  extract only the basic labels (``colors'') encoding the communications. 
\end{itemize}
Of course, humans are still humans, not social atoms; thus, in contrast to quantum physics, it is impossible to create human gain mediums
composed of completely indistinguishable creatures.  People still have say names, gender, nationality, but such their characteristics are ignored in the regime of social lasing. 

One of the basic components of lasers, both physical and social, is a resonator \cite{L4}. It plays the double role: 
\begin{itemize}
\item amplification of the beam (of physical vs. information) radiation;
\item improving coherence of this beam.
\end{itemize}
Social laser resonators play the crucial role in generation of coherent information waves of high amplitude. They are established 
via the {\it internet-based  Echo Chambers} associated with social networks, blogs, and YouTube channels. Their functioning is based on the feedback
process  of posting and commenting, the process that exponentially amplifies the information waves that are initially induced by 
mass-media.  Echo Chambers improve the coherence of the information flow through the statistical elimination of communications that 
do not match the main stream. This statistical elimination is a consequence of the bosonic nature of the quantum information field
(sections \ref{BF}, \ref{BFA}).
Although this quantum process of coherence generation dominates in Echo Chambers, we should not ignore other technicalities increasing coherence (sections \ref{CO}, \ref{G}), such as censorship of moderators and the dynamical evaluation system of search engines of, e.g., Google, YouTube, or Yandex. The latter system elevates approachability of posts, comments, and videos depending on the history of their reading (seeing) and reactions to them say in the form of new comments.   

This a good place to recall that the quantum-like Hilbert space formalism if widely used for modeling of information processing 
by Internet search engines; in particular, for information retrieval  \cite{van0}-\cite{Me1}. 
 
We compare functioning of the optical and information mirrors (section \ref{C1}). The latter represents the feedback process in the internet systems such as, e.g., YouTube. In contrast to the optical mirror, the information mirror not only reflects excitations of the quantum information field, but also multiplies them. Thus, this is a kind of reflector-multiplier (section \ref{C2}). As the result of this multiplication effect, social resonators are more effective than physical ones. However, as in physics, resonator's efficiency 
depends on a variety of parameters. One of such parameters is the coefficient of reflection-multiplication (section \ref{RM}).
We analyze the multilayer  structure of an information mirror and dependence of this coefficient on the layer (section \ref{RM}). 

The main output of this paper is presented in section \ref{TSU} describing the quantum-like mechanism of generation of big waves of coherent information excitations.  

We start the paper with compact recollection of the basics of the social laser theory distilled from technical details and mathematical 
formulas. We present the basic notions of this theory such as social energy  (section \ref{SE})) and social atom, human gain medium
(section \ref{SA})), information field (section \ref{IF})), the energy levels structure of social atoms (section \ref{EL}), 
spontaneous and stimulated emission of information excitations (section \ref{SPS}). Finally, we conclude the introduction by the schematic presentation of functioning of  social laser (section \ref{SLS}). The role of information overload in approaching indistinguishability of information communications, up to their basic labels, {\it quasi-colors}, is discussed in section \ref{IO}.

\section{Basics of lasing}

\subsection{Physical laser}

The basic component of physical laser is a {\it gain medium}, an ensemble of
atoms. Energy is pumped into this medium aimed to approach the state of
{\it population inversion}, i.e, the state where more than 50\% of atoms are
excited \cite{HX} . Then a coherent bunch of photons is injected into the gain medium
and this bunch stimulates the {\it cascade process of emission} of the coherent
photon-beam. If the power of pumping is very high, i.e., it is higher than
the so-called lasing threshold, all energy of pumping is transferred into
the output beam of coherent radiation. To make this beam essentially
stronger, the laser is equipped by an additional component, {\it laser's
resonator} (typically in the form of optical cavity). Laser's resonator
also improves coherence of the output beam, by eliminating from the beam,
photons that were generated via spontaneous emission in the gain medium \cite{HX}.

Typically in physics coherence is formulated in physical waves terms, as electromagnetic 
waves going in phase with the same direction of propagation and frequency. For us it is convenient, 
to reformulate this notion by excluding any reference to waves in the physical space, since we want to
move to the information space. Instead of the wave picture we can use the photon picture, so 
a propagating wave is represented as a cloud of energy quanta.\footnote{This is the Fock representation in quantum 
field theory.}  Coherence means that they have the 
same energy (frequency) and the direction of propagation - photon's wave vector. We remark that a photon also has 
additional characteristics such as polarization, the quantum version of the ordinary polarization 
of light. For convenience of further considerations, let us call all characteristics of a photon additional to its energy 
{\it quasi-color.} We recall that the usual light's color is determined by photon's energy (frequency). 
So, a photon has its color and quasi-color.   

\subsection{Social energy}
\label{SE}

The notion of the social energy is the main novel component of our quantum-like
modeling. To justify the use of a social analog of the physical energy, we use
the quantum-mechanical interpretation of energy, not as an internal feature
of a system, but as an observable quantity. Thus, like in the case of an
electron, we cannot assign to a human the concrete value of the
social energy. There are mental states in superposition of a few different
values of the social energy. However, by designing proper measurement procedures
we can measure human's energy, see \cite{L1,L4} for details. 

The social energy is a special form of the psychic energy. We recall that at the end of 19th - beginning of 20th 
century psychology was strongly influenced by physics, classical statistical physics and thermodynamics (in works of 
James and Freud), later by quantum physics (in works of Jung). In particular, the leading psychologists of that time 
have actively operated with the notion of psychic energy \cite{J}-\cite{PJ}. Later psychologists essentially lost the interest to construction of general theories and, in particular, operating with the notion of the social energy.
 
Recently, the notion of social energy attracted a lot interest in  economics and finances,  multi-agent modeling, evolution theory and
industrial dynamics  \cite{SEM1,SEM2,SEM3}. Of course, these novel as well as old (Freud-Jung) studies are supporting for our model.
However, we emphasize that the application of the quantum  (Copenhagen) methodology simplifies and clarifies essentially the issue of the social energy. We treat it operationally as an observable on a system, a human being. In contrast to say Freud, we are not interested 
in psychic and neurophysiological processes of generation of psychic energy (see appendix for a brief discussion).   

\subsection{Human gain medium, social atoms}
\label{SA}

The basic component of social laser is a {\it gain medium}, an ensemble of people. As was already mentioned, to initiate 
lasing such a gain medium should consist of {\it indistinguishable} people so to say without tribe, without cultural, national, religious, and ideally sex differences. Such beings  are called social atoms. (It is not clear whether they still can be called humans?) Of course, 
people still have aforementioned characteristics, in some contexts they remember that they are say men or women, or even 
christian, or Swedish. We discuss contexts in which people behave as indistinguishable, as social atoms. Creation of  such behavioral contexts is the first step towards initiation of social lasing.    

\subsection{Information field}
\label{IF}

We recall that in quantum physics the electromagnetic field is treated as a carrier of interactions. In the quantum framework,
interaction cannot be represented as it was done classically, by force-functions. Quantum interaction is of the information nature.
In quantum information theory, excitations of the quantum electromagnetic field, photons, are carriers of information.
At the same time each excitation also carries quantum of energy. 

This quantum picture is very useful for general modeling of information fields generated by mass-media and Internet. 
Communications emitted by newspapers, journals, TV, social networks, blogs are modeled as excitations of a quantum 
information field, as quanta of information and social energy. 

As we know, the quantum description is operational, this is only the mathematical symbolism used for prediction of probabilities.
Even the quantum electromagnetic field cannot be imagined as a ``real wave'' propagating in space-time.\footnote{In the formalism, 
this is a distribution, generalized function, with operator values. Hence, this is  a very abstract mathematical structure. It 
is useful for accounting of the numbers of energy quanta and description of the processes of their emission and absorption.}
 On one hand, this impossibility of visualization is disadvantage 
of the quantum description comparing with the classical one \footnote{We remark that visualization of the classical electromagnetic field is neither so straightforwards as it might be imagined. The electromagnetic waves were invented as the waves propagating in the special media, aether, similarly to acoustic wave propagating in air. Later Einstein removed aether from physics. The picture of a vibrating medium became inapplicable. So, electromagnetic waves are vibrations of vacuum. This is not so natural picture for visualization of this process.}. On the other hand, this is the great advantage, since it provides the possibility for generalizations having no connection with the physical space-time. 

Thus, we model information field as a quantum field with communications (generated, e.g., by mass-media) as quanta carrying social energy and some additional characteristics related to communication's content. As was already emphasized, quantum description 
is applicable to fields with indistinguishable excitations, where indistinguishability is considered up to observable 
characteristics. And ``observable'' means those characteristics that people assign to communications.  These are labels of 
communications, say ``terrorism'', ``war in Syria'', ``corona-virus'' and so on. Such labels we shall call {\it quasi-colors of information excitations}, these are analogs 
of photon's wave vector and polarization. Thus, each communication is endowed with   
a quasi-color.  It also carries a quantum of energy, its value we consider as communication's color.
Thus allegorically we can speak about red, blue, or violate information.  

{\it Content-ignorance} (up to communication's quasi-color and color)  is the crucial feature of  applicability of the quantum formalism. 

\subsection{Information overload}
\label{IO}

Why do social atoms compress contents of communications to quasi-colors? The most important is information overload. The information flows generated by mass-media and Internet are so powerful, that people are not able to analyze communication's content deeply, they just scan its quasi-color and absorb quantum of the social energy carried by this communication. They simply do not have computational and 
time resources for such an analysis. It is also crucial that people lose their identity, so they become social atoms. For a social 
atom, there are no reasons, say cultural or religious, to analyze news, he is fine with just absorption of labels (quasi-color)  and social energy (color) assigned to them.  

\subsection{Energy levels of social atoms}
\label{EL}

Consider for simplicity social atoms with just two energy levels, excited and relaxed, $E_1$ and $E_0.$ The difference between these
levels, 
\begin{equation}
\label{pe}
E_{\rm{a}}=E_1 - E_0,
\end{equation}  
is the basic parameter of a social atom, its color.  A social atom reacts only to a communications carrying energy $E_{\rm{c}}$ matching 
his color:
\begin{equation}
\label{pe7}
E_{\rm{a}} =E_{\rm{c}}.
\end{equation}  

If a communication carries too high energy charge, $E_{\rm{c}}$ larger than  $E_{\rm{a}}$ (``a social  atom is yellow, but a communication is blue''), then an atom would not be able to absorb it. Say a communication carrying social energy $E_{\rm{c}}$  is a call  for upraise against the government. And an atom is a bank clerk in Moscow, who has the 
liberal views and hates the regime, but the energy  of his excited state is too small to react to this call. If $E_{\rm{c}}$ is less than $E_{\rm{a}}$
(``an atom is blue, but a communication is  yellow''),  then an atom would not be excited by this communication. The communication would be simply ignored. As well as a physical atom, a social atom cannot collect social 
energy continuously from communications carrying small portions of energy (comparing with $E_{\rm{a}}=E_1- E_0),$ it either absorbs communication (if the colors of an atom and communication match each other) or it does not pay attention to it. In the same way, a social atom cannot ``eat'' just a portion of energy carried by too highly charged 
communication. 

In physic's textbooks, the condition of absorption of energy quantum by atom is written as the precise equality:
\begin{equation}
\label{pep}
E_{\rm{photon}} = E_{\rm{a}}.
\end{equation}
However, precise equalities  are only mathematical idealizations of the real situation. Photon-absorption condition  (\ref{pep}) is satisfied only approximately:
\begin{equation}
\label{pe1}
E_{\rm{photon}} \approx E_{\rm{a}}.
\end{equation}  
The spectral line broadening is always present. The difference between the energies of atom levels is the mean value (average) of the Gaussian distribution, a bell centered at this point of the energy axis. The dispersion of the Gaussian distribution depends on an ensemble of  atoms. Ensembles with small dispersion 
are better as gain mediums for lasing, but deviations from exact law (\ref{pep}) are possible.

It is natural to assume Gaussian distribution realization of exact laws even for social systems; in particular, absorption of  
of excitations of the quantum information field by social atoms. Thus deviations from (\ref{pe}) are possible. But, a good 
human gain medium should be energetic homogeneous. Therefore, the corresponding Gaussian distribution should has very small 
dispersion. 
 
\subsection{Shock-news as the best source of energy pumping}
 
Shock-news, say a catastrophe, war, killed people, epidemy,  terror attack, are very good for energy pumping to a social gain medium.
The modern West-society is characterized by the high degree of excitation, the energy $E_1$ of the excited level  is sufficiently high -
otherwise one would not be able to survive: life in megalopolis,  long distances, high intensity of the working day, and so on. On the other hand, the energy $E_0$  of the relaxation level is very low - one who is  living on the state-support, say in Sweden, has practically zero excitement, often his state is depressive. Hence,   $E_{\rm{a}} = E_1- E_0 $ is high and a social atom would absorb only communications carrying 
very high energy: as in aforementioned shock-news or  say in TV-shows people should cry loudly, express highly emotional psychic states.
 Since $E_{\rm{a}}$ is high (say blue), people would not pay attention on so to say plane news (say red colored). Even scientific news attract attention only if they are very energetic, carry big emotional charge (blue or even better violate). 
  
 However, shock-news are very good for energy pumping not only because they carry a high charge of social energy, but also because they 
are very good to peel communications from content. Labels (quasi-colors) such as say ``coronavirus is bio-weapon'' lead to
immediate absorption of communications, social atoms react immediately to the instinctive feeling of danger.

\subsection{Spontaneous and stimulated emission}
\label{SPS}

In our quantum-like model (similarly to physical atoms),  social atoms can  both absorb and emit quanta of the social energy. As in physics, there are two types of emission, spontaneous and stimulated.
  
The spontaneous emission happens without external interaction, a social atom spontaneously emits a quantum of social energy, in the form of some social action. Such spontaneous actions are not coherent, different atoms 
do different things, quasi-colors of social energy quanta emitted spontaneously can be totally different.  Such emissions generate a {\it social noise} in a human media, noise that is unwanted in social lasing. In particular, spontaneous emission noise disturbs functioning 
of internet echo-chambers. 

On the other hand, emission of quanta of social energy can be stimulated by excitations of the information field. In the very simplified picture, it looks so. An excited social atom by interacting with an information excitation emits (with some probability) quantum of social energy. The most important feature of this process is that the quasi-color of the emitted quantum coincides with the quasi-color of stimulating communication. This is the root of the coherence in output beam of lasers, both social and physical. (The colors also coincide, see section \ref{EL}). 

\subsection{Quantum information field as a bosonic field}
\label{BF}

In reality, the process of stimulated emission is more complicated. It is important that the information field  (similarly to the quantum electromagnetic field) satisfies Bose-Einstein statistics. This is a thermodynamical  consequence  \cite{L1} of indistinguishability of excitations: two excitations with the same social energy and quasi-color are indistinguishable.
As was shown in  \cite{L1}, by using the Gibbs' approach based on consideration of virtual ensembles of  indistinguishable systems (or any origin) we obtain the standard quantum classification of possible statistics, Bose-Einstein, Fermi-Dirac, and parastatistics. Indistinguishability is up to energy (for the fixed quasi-color). Hence, by taking into account that the number of communications carrying the same charge of social energy can be arbitrary, we derive the Bose-Einstein statistics for the quantum information field (see  \cite{L1} for derivation's details). 

 Interaction of atomic-like structures with bosonic fields  are characterized by the following property: probability of stimulated emission from an atom increases very quickly with increasing of the power of a bosonic field. An excited social atom reacts rather weakly to the presence of  a few information excitations. But, if they are many, then it cannot stay indifferent. In fact, this is just a socio-physical expression of the well known {\it bandwagon effect} in humans' behavior \cite{BEF}. In contrast to psychology, we are able to provide the mathematical field-theoretical model for such an effect. 

We consider the fixed energy (frequency) mode of the  quantum electromagnetic field. For  fixed quasi-color mode $\alpha,$ $n$-photon state $\vert n, \alpha\rangle,$  
can be represented in the form of the action of the photon creation operator $a_\alpha^\star$ corresponding to this mode on the vacuum state $\vert 0\rangle$: 
\begin{equation}
\label{CR1}                                                                 
\vert n, \alpha\rangle =[(a_\alpha^\star)^n/ \sqrt{n!}] \vert 0\rangle
\end{equation}                                                            
This representation gives the possibility to find that the transition probability amplitude from the state $\vert n, \alpha\rangle$ to the state $\vert n+1, \alpha\rangle$ 
equals to $\sqrt{(n+1)}.$ On the other hand, it is well known that the reverse process of absorption characterized by 
the transition probability amplitude from the state $\vert n, \alpha\rangle$ to the state $\vert (n-1), \alpha \rangle$ equals to $\sqrt{n}.$ . Generally, for a quantum bosonic field increasing  the number of its quanta  leads to increasing the probability of generation of 
one more quantum in the same state. This constitutes one of the basic quantum advantages of laser stimulated emission showing that the 
emission of a coherent photon is more probable than the absorption. 

Since, as shown in \cite{L1}, indistinguishability, up to energy (color)  and quasi-color, of information excitations leads to the Bose-Einstein statistics, we can use the quantum operational calculus for bosonic fields even for the quantum information field and formalize in this way the bandwagon effect in psychology \cite{L5}.

\subsection{Social action terminology}

This is the good place to recall that in our considerations the notion ``social action'' is treated very widely, from a purely information action, as posting a communication at Face Book or commenting one of already posted communications, to a real physical action, as  participating in a demonstration against Putin or Trump, or supporting  government's policy on ``self-isolation''.  The previous works on social laser \cite{L1}-\cite{L4} emphasized external representation of social actions, say in the well known color revolutions. In this paper, we are more interested in their representation in information spaces, e.g., spaces of social networks.
But, we are even more interested    in internal representation of some social actions as decision makings. And a decision can have different forms,
not only ``to do''-decisions, but also ``not to do''-decisions. The decisions of the latter type also consume energy and social atoms 
transit from the excited state to the relaxed one. 

It is also important to point to the unconscious character of many (or may be majority) of our decisions. For example, people can support 
(or not support) societal policies totally unconsciously. To make such decisions, they consume social energy.

\section{Social lasing schematically}
\label{SLS}

Mass-media and internet pump social energy into a gain medium composed of social atoms to approach the population inversion - to 
transfer the majority of atoms into excited states. Then a bunch of communications of the the same quasi-color and energy (color) 
matching with the resonant energy of social atoms is injected in the gain medium. In the simplified picture, each communication  
stimulates a social atom to emit a quantum of social energy having the same quasi-color with its stimulator. Resulting two excitations stimulate two social atoms to emit two quanta, the latter two quanta generate four and so on, after say 20 steps there are $2^{20},$
approximately one million of information excitations of the same (quasi-)color. In reality, the process is probabilistic: an atom 
reacts to stimulating information excitation only with some probability. The later increases rapidly with increasing of the density of 
 the quantum information field. 

Now, we discuss the basic counterparts of social lasing in more detail:

\begin{itemize}
\item Each  information communication carries a quantum of  social energy. The corresponding mathematical model is of the quantum field type, the information field. Quanta of social energy are its excitations.
\item Each social atom is characterized by the  social energy spectrum; in the simplest case of two levels, this is the difference between the energies of the excitation and relaxation states, $E_{\rm{a}}=E_1 - E_0.$
\item Besides of social energy, the excitations of the information field are characterized by other labels, quasi-colors. 
\item Coherence corresponds to social color sharpness; ideal social laser emits a single mode of quasi-color, denoted say by the symbol $\alpha.$
\item Humans in the excited state interacting with $\alpha$-colored excitations of the information field also emit $\alpha$-colored excitations. 
\item The amount of the social energy carried by communications stimulating lasing should match with  resonance energy $E_{\rm{a}}$ of social atoms in the human gain medium.  
\item To approach the population inversion, the social energy is pumped into the gain medium. 
\item This energy pumping is generated by the mass-media and the Internet 
sources. 
\item The gain medium should be homogeneous with respect to the social energy spectrum. In the ideal case, all social atoms in the gain medium should have  the same spectrum, $E_{\rm{a}}.$ However, in reality, it is impossible to create such a human gain medium. As in physics, the {\it spectral line broadening} has to be taken into account.
\item Social quasi-colors of excitations in the energy pumping beam have no straightforward connection  with the quasi-color of  excitations in the output beam generated by stimulating emission. 
\item Information excitations follow the Bose-Einstein statistics.
\item This statistics matches with the Bandwagon effect in psychology.
\item  The probability of emission of the $\alpha$-colored excitation by a social atom in a human gain medium
 increases very quickly  with the increase of the intensity of the information field on the $\alpha$-colored mode. 
\item This behavior generates the cascade of coherent social actions corresponding to
information excitations emitted by social atoms. 
\end{itemize}
 
For example, a gain medium consisting of humans in the excited state  and stimulated by the anti-corruption colored information field would  ``radiate'' a wave of anti-corruption protests. 
The same gain medium stimulated by an information field carrying another social color would generate the wave of actions corresponding this last color.

\section{Echo Chambers as social resonators}

The general theory of resonators for social lasers is presented in \cite{L4}. Here we shall consider in more detail special, but at 
the same very important  type of social resonators, namely, Internet based {\it Echo Chambers}. We recall that  an Echo Chamber is a system in that some ideas
and behavioral patterns are amplified and sharped through their feedback propagation inside this system. In parallel to such amplification, communications carrying (as quasi-color)  ideas and behavioral patters different from those determined 
by the concrete Echo Chamber are suppressed.  

In our terms, an {\it  Echo Chamber is a device for transmission and reflection of excitations of the quantum information field.} Its main  purpose is amplification of this field and increasing its coherence via distilling from  ``social noise''. The latter function
will be discussed later in more detail. The Echo Chamber is also characterized by the resonance  social energy $E_{\rm{a}}$ of its social atoms. For simplicity, it is assumed that all social atoms have the same resonance energy   
$E_{\rm{a}}.$ (In reality, resonance energy of social atoms is a Gaussian random variable with mean value $E_{\rm{a}}.)$      

We underline that in this paper an Echo Chamber is considered as a component of  the social laser, its resonator. Comparing with physics we can say that this is an analog of an optical cavity of the physical laser, not optical cavity by itself. The coherent output of an Echo Chamber, the quasi-color of this output, is determined not only by the internal characteristics of the Echo Chamber, but also by the quasi-color of stimulating emission.    
 
\subsection{Echo Chamber  based on Internet social group}

Let us consider functioning of some Internet-based Echo Chamber; for example, one that is based on some social group in Face Book 
(or its Russian version ``Vkontakte'') and composed of social atoms. The degree of their indistinguishability can vary depending on the concrete Echo Chamber. Say, names are still present in {\it Face Book}, but they have some meaning only for the restricted circle of friends; in {\it Instagram} or {\it Snapchat}, even names disappear and social atoms operate just with nicknames. 
By a social group we understand some sub-network of say Face Book, for example, social group ``Quantum Physics''.  The main feature 
of a social group  is that all  posts and comments are visible for all members of this social group. Thus, if one from the group puts 
a post, then it would be visible for all members of this social group, and they 
would be able to put their own comments or  posts related  to my initiation post. This is simplification of the general structure of posting in  Face Book, with constraints that are set by clustering into ``friends'' and ``followers''.      

\subsection{Comparing optical and information mirrors}
\label{C1}

We assume that the ensemble of social atoms of this Echo Chamber approached population inversion, so the majority of them are already excited. A bunch of communications of the same quasi-color $\alpha$ and carrying quanta of social energy $E_{\rm{c}} = E_{\rm{a}}$ is injected in the Echo Chamber. Excited social atoms interact with the stimulating communications and emit (with some probability) information excitations of the same quasi-color as the injected stimulators.  These emitted quanta of social energy are represented in the form of new posts in Echo Chamber's social group. Each post plays the role of a mirror, it reflects the information excitation that has generated this post. 

However, the analogy with the optics is not straightforward. In classical optics, each light ray is reflected by a
mirror again as  one ray. In quantum optics, each photon reflected by a mirror is again just one photon. An ideal mirror reflects all photons (the real one absorbs some of them). 

In contrast, ``the mirror of an Echo Chamber'', the information mirror,  is  a {\it multiplier.} A physical 
analog of such a multiplier mirror would work in the following way. Each light ray is reflected as a bunch of rays or in the quantum picture (matching better the situation),  each photon by interacting with such a mirror generates a bunch of photons. Of course, the usual physical mirror cannot reflect more photons than the number of incoming ones, due to the energy conservation  law. Hence, the discussed device is hypothetical.

This is a good place to remark that, as was mentioned, a photon should not be imagine as a metal ball reflecting from mirror's surface. A photon interacts with the macro-system, the mirror, and the latter emits a new photon that is identical to the incoming one,  up to the direction of spatial propagation. It seems to be possible to create a kind of a mirror with the complex internal structure (composed of special materials) such that it would generate emission of a bunch of photons. Of course, such a multiplier mirror cannot function without the energy supply.

\subsection{Information mirror as multiplier}
\label{C2}

The Internet-based system of posting news and communications works as  a multiplier mirror. Each posted news or communication
emits a bunch of ``information rays'' directed to all possible receivers -  the social atoms of Echo Chamber's social group. In the quantum model, each post works as an information analog of photon's emitter. It emits quanta of social energy, the power of the information field increases. Consequently excited social atoms emit their own posts and comments with higher probability. We repeat that new posts have the same quasi-color as the initiating information excitations that were injected in the Echo Chamber.

It is also important to remind that the process of stimulated emission is probabilistic. Members of the social group would react to
newly posted message only with some probability. And resulting from the bononic nature of the quantum information field, this probability increases rapidly with increasing of  field's power.

By {\it reaction we understood emission of a new message}, say a comment. If a social atom simply reads a posted communication, but does not emit its own  information excitation, then we do not consider such reading as a reaction. For the moment, we consider only the  process of stimulated emission. Later we shall consider absorption as well. In the 
latter, reaction means transition from the ground state to the excited state; so, not simply reading. (In principle, a relaxed 
atom can read a post or a comment without absorbing a quantum of social energy  sufficient for approaching the state of excitement.)

The crucial difference from physics is an apparent violation of the energy conservation law (see appendix for the discussion on this question). Each post in a social group
works as a social energy multiplier. Thus information excitations in the Echo Chamber generated by posted communications 
not only increase the probability of emission of new information excitations by excited atoms, but they also perform
the function of additional energy pumping into the gain medium (social group). 
Relaxed social atoms can absorb social energy not only from 
externally pumped messages from mass-media, TV and other social networks, but even from their own Echo Chamber. Then they also 
emit new posts and so on.

\subsection{The role of indistinguishability}  
\label{BFA}  

The main distinguishing feature of the quantum information field is its bosonic nature. We now emphasize the impact  of the  
bosonic structure  to coherence of the information field inside of an Echo Chamber.
As was already noted (section \ref{BF}), the interaction of a social atom with the surrounding bosonic field depends crucially 
on the power of this field, the probability of emission of energy quantum by an excited social atom increases very quickly 
with increasing of field's power. 

Now, we stress again that a social atom (as well as a physical atom) distinguishes the modes 
of the field corresponding to different quasi-colors. The probability of emission of a quantum of the fixed quasi-color $\alpha$ 
depends on the power of the field's mode colored by $\alpha.$ Thus, if the power of the $\alpha$-mode essentially higher than the power 
of the mode colored by $\beta,$ then with very high probability social atoms would emit $\alpha$-colored energy quanta (in the form of posts, comments, and videos). Social atoms would ignore the $\beta$-colored energy quanta, the probability of emission of such 
quantum (and hence the increase of the power of the $\beta$-mode)  is practically zero. If a social atom emits a communication, 
colored by  $\beta,$  then this information excitation would not attract attention of social atoms who are busy with communications colored by $\alpha.$ 

As was already emphasized, the crucial role is played by indistinguishability, up to the quasi-colors, of the excitations of the information field. Social atoms should process information in the regime of label scanning, without analyzing its  content. As was discussed, the easiest way to establish the indistinguishability regime of information processing is to generate an information overload in the gain medium composed of social atoms. Of course, the lose of individuality by social atoms is also very important, people 
``without tribe'' are better accommodated to perceive information in the label-scanning regime. In this regime, one would never absorb the main information of the $\beta$-labeled communication, say statistical data. 

In this section, we considered the quantum-like nature of coherence of the information waves generated in Echo Chambers. Thisis
indistinguishability of information excitations, the label-scanning regime. The information overload and the loss of individuality by social atoms are the main socio-psychological factors leading to this regime.  

In following sections \ref{CO}, {G}, we consider supplementary factors increasing information field's coherence.              

\section{Generation of coherence information waves in society}
\label{TSU}

Now, we connect a social resonator, e.g., in the form of an internet-based Echo-Chamber, to the social laser device 
described in section \ref{SLS}. As the result of the feedback processing of information in the Echo Chamber, the power 
and coherence of the information field increases enormously. 

One of the ways to  consume the huge energy of this information field is
to realize it in the form of physical social actions, mass protests, e.g., demonstrations or even a wave of violence. This is the 
main mechanism of color revolutions and other social tsunamis \cite{L1}-\cite{L4}. 

However, in this paper we are more interested 
in the purely information consumption of the social energy of the coherent information field prepared in an Echo Chamber, 
namely, for internal decision making. Decision making on a question that important for society
is also a social action; in particular, it consumes social energy. Now, suppose that say a government needs some coherent 
and rational (from its viewpoint) decision on some question. It can use a powerful social laser. This is a good place to remark 
that an ensemble of Echo Chambers can be used coherently with stimulation by the same quasi-color $\alpha$  corresponding to the desired decision. By emitting the information excitation a social atom confirms his-her support of the $\alpha$-decision. Such social action is realized 
in the mental space of social atoms, but, of course, it has consequences even for associated actions in the physical space. 

If the  wave in the information space generated by a powerful social laser can approach the steady state, then social atoms 
live in the regime of the repeated confirmation of the internal $\alpha$-decision: an atom emits  and relaxes, then 
he-she again absorbs another  $\alpha$-excitation  and moves to the state of excitement and so on. In this situation of surrounding by 
the information field of huge power concentrated on the same  $\alpha$-mode,  the colors of the energy pumping and stimulated 
emission coincide.  Such repeating of the same $\alpha$-decision is similar to concentration on the idea-fix and can lead to 
the state of psychosis  and panic (see Freud \cite{}).

\section{Technicalities}

\subsection{Losses, coefficient of reflection-multiplication}
\label{RM}

As in physical lasing, the above ideal scheme is complicated by a few factors related to losses of social energy 
in the Echo Chamber. As is known, not all photons are reflected by mirrors of the optical cavity, a part of them is absorbed by 
the mirrors. The coefficient of reflection plays the fundamental role. The same problem arises in social lasing. An essential 
part of posts is absorbed by the information mirror of the Echo Chamber: for some posts, the probability that they would be read by members of the social group  is practically zero. Additional (essential) loss of social energy is resulted from getting rid  of
 communications carrying quasi-colors different from the quasi-color $\alpha$ of the bunch of the communications initiating  the 
feedback dynamics in the Echo Chamber. Such communications are generated by spontaneous emission of atoms in the social group. 

The real model is even more complex. The information mirror is not homogeneous, ``areas of its surface'' differ by the degree 
of readability and reaction. The areas can be either rigidly incorporated in the structure of the social group or be  formed in the process of its functioning. 

For example,  ``Quantum Physics'' group has a few layers that are rigidly incorporated in its structure.
One of them is ``Foundations and Interpretations''. This sublayer of the information mirror ``Quantum Physics'' has rather low visibility, due to a variety of reasons. Once, I posted in ``Quantum Physics'' a discussion on quantum information and quantum nonlocality. And I discovered that the social group moderators control  rigidly the layer structure. 
The message that my post should be immediately moved to this very special area of the information mirror, ``Foundations and Interpretations'', approached me in a few minutes. It looks that even in such a politically neutral social group moderators work in the online regime.

As an example of functionally created information layers, we can point to ones which are coupled to the  names of some members of the social group, say ``area'' related to the posts of a Nobel Prize Laureate has a high degree of readability and reaction. But, of course, one need not be such a big name to approach a high level of  readability and reaction. For example, even in science the strategy of
active following to the main stream speculations can  have a very good effect. Top bloggers and youtubers create areas of the information mirror with high coefficients of reflection-multiplication (see below (\ref{LI1})) through collecting subscriptions to their blogs and YouTube channels.      

It is clear that the probability of readability and reaction to a post depends heavily on the area of its location in the information space  of a social group or generally Face Book, YouTube, or Instagram. The reflection-multiplication  coefficient of the  information mirror varies essentially. 
 
Consider first the physical mirror and photons reflected by it. From the very beginning, it is convenient to consider an  inhomogeneous mirror with the reflection coefficient depending on mirror's layers.
Suppose that $k$-photons are emitted to  area $x$ and $n$ of them were reflected, i.e., $(k-n)$ were absorbed. 
Then the probability of reflection by this area 
\begin{equation}
\label{LI}
P(x) \approx n/k, \mbox{for large} \;  k. 
\end{equation}
Now, for the information mirror, consider a sequence of posts, $j=1,2,..., k,$ that were put in its area $x.$
Let $n_j$ denotes the number of group's members who reacts to post $j.$ Each $n_j$ varies between $0$ and $N,$ 
where $N$ is the total number of group's members. Then coefficient of reflection-multiplication 
\begin{equation}
\label{LI1}
P(x) \approx (\sum_j^k n_j)/ kN, \mbox{for large} \;  k, N.
\end{equation}
 If practically all posts generate 
reactions of practically all members of the group, then $n_j\approx N$ and $P(x)\approx 1.$  

\subsection{Improvement of coherence: direct and indirect filtering}
\label{CO}

We have already discussed in detail the multilayer structure of the information mirror of an Echo Chamber. 
This is one of the basic information structures giving the possibility to generate inside it the information field
of the very high degree of coherence: a very big wave of information excitations of the same quasi-color, the quasi-color 
of stimulating communications. It is sufficient to stimulate atoms having the potential of posting in the areas of the information surface having the high coefficients of reflection-multiplication. These areas would generated a huge information wave dirrected 
to the rest of the social group. 

Spontaneously emitted communications would be directed to areas with the low coefficients of reflection-multiplication.
How is this process directed by the Internet engines? It is described by the model of {\it the dynamical evaluation 
of the readability history of a post}. We shall turn to this model in section \ref{G}. 

Although the dynamical evaluation plays the crucial role in generating the coherent information waves, one has not to ignore the impact of straightforward filtering. We again use the analogy with physics. In the process of lasing, the dynamical feed back process in the cavity excludes the excitation of the electromagnetic field propagating in the wrong directions. In this way, laser generates the sharply 
directed beam of light. However, one may want some additional specification for excitations in the light beam. For example, 
one wants that all photons would be of the same polarization. It can be  easily done by putting the additional filter, the polarization filter, that eliminates from the beam all photons with ``wrong polarization''. Of course, the use of an additional filter would weaker 
the power of the output beam. The latter is the price for   coherence increasing. In social lasing, the role of such polarization
filters is played by say Google, Face Book, Instagram, or   Yandex control filtering, e.g., with respect to the political correctness constraints. Besides numerous moderators, this filtering system uses the keywords search engines  as well as the rigid system of ``self-control''.  In the latter, users report on ``wrongly colored posts and comments'' of each other; the reports are directed  both to the provider and to social groups - to attract the attention to such posts and comments.

\subsection{Dynamical evaluation system}
\label{G}
  
The dynamical evaluation system used, e.g., by YouTube, increases post's visibility on the basis of its reading history, more readings imply higher visibility (at least theoretically). However, the multilayer structure of the information mirror of YouTube should also be taken into account. 

The main internet-platforms assign  high visibility to biggest actors of the mass-media, say BBC, EuroNews, RT,  that started  to use actively these platforms. Then, and this is may be even more important, these internet-platforms assigns high visibility to the most popular topics, say nowadays the coronavirus epidemy, videos, posts,  and comments  carrying this quasi-color are elevated automatically in the information mirrors of Google, YouTube, or Yandex.

Of course, the real evaluation system of the main internet actors is more complicated and the 
aforementioned dynamical evaluation system is only one of its components, may be very important. 
We would never get the answer to the question so widely discussed in communities of bloggers and youtubers:
How are the claims on unfair policy of the internet-platforms justified? By unfair policy they understand 
assigning additional readings and likes to some internet-communications or withdraw 
some of them from other communications.\footnote{I can only appeal to my own rather unusual experience from the science field.
Once, I was a guest editor of a special issue (so a collection of papers about some topic).
In particular, my own paper was published in the same issue. This is the open access journal of v top ranking, a 
part of Nature publishing group. Nowdays, all open access journals qualify papers by the number of downloads and readings.
(So, this is a kind of YouTubing of science.) My paper was rather highly estimated in these numbers. But, suddenly 
I got email from the editors that, since I put so much efforts to prepare this issue, I shall got as a gift additional 1200 
downloads. Of course, I was surprised, but I did not act in any way and really received this virtual gift... After this event,
I am very suspicious to numbers of downloads and readings that I can see in the various internet systems. If such unfair behavior 
is possible  even in science, then one can suspect that it is really not unusual.}

\section{Concluding remarks}  

Starting with presentation of the basics of social lasing, we concentrated on functioning of one of the most important kinds of 
social resonators, namely internet based Echo Chambers. We analyzed similarities and dissimilarities of optical and 
information mirrors. The main distinguishing feature of the latter is its ability not only reflect excitations of 
the quantum information field, but also multiply them in number. The coefficient of reflection-multiplication is the basic 
characteristic of the information mirror. We point to the layer structure of the information mirror of an Echo Chamber;   the coefficient of reflection-multiplication varies depending on mirror's layer. 

We emphasized the bosonic nature of the quantum information field. This is a straightforward thermodynamical consequence \cite{} of indistinguishability of information excitations, up their quasi-colors. Being bosonic, the information field increases tremendously 
the speed and coherence of stimulated emission of information excitations by excited social atoms. 

Social atoms, ``creatures without tribe'', form the gain medium of a social laser. In contrast to quantum physics, we cannot 
treat real humans as totally indistinguishable. This is a good place to remind once again that our social lasing as well as generally decision making modeling  is {\it quantum-like}. Quantum features are satisfied only approximately. This point is often missed 
in presentation of ``quantum models'' for cognition and decision making.

In sections \ref{CO}, \ref{G}, we discuss some technicalities related to functioning of internet based social groups and generally Google and YouTube. This discussion plays only a supplementary role for this paper. It would be fruitful to continue it and especially 
to discuss exploring of the quantum-like features of users and information supplied to them by the Internet (cf., for example, 
with studies on quantum-like modeling of information retrieval \cite{}).  

In appendix, we discussed very briefly interrelation between the psychic energy and the physical energy of cells' metabolism. It is very important to continue this study in cooperation with psychologists and neurophysiologists. 

The main output of this paper is description of the mechanism of generation of big waves of coherent information carrying  
huge social energy, a kind of information tsunamis. We especially emphasize listing of the basic conditions 
on  human gain media and the information field generated by mass media and amplified in Echo Chambers leading to successful 
generation fo such waves. 

The author recognizes very well that this study is still one of the first steps toward well 
elaborated theory. 

\section*{Appendix: The law of conservation of psychic energy}

Above we wrote about an ``apparent violation'' of the law of conservation of the social energy. We briefly discuss this point.
The social energy is the special form of the psychic energy. Hence, by discussing the conservation law we cannot restrict 
consideration solely to the social energy. The detailed analysis of transformation of different forms of the psychic energy 
and its origin in neurophysiological processes and finally the physical energy  generated by cells' metabolism  was 
presented by Freud \cite{}. We do not plan to discuss here the Freud's hydrodynamical model for psychic energy 
transformations. We want to elevate the crucial difference of the energy transfer from the information field to social atoms 
from the energy transfer from the electromagnetic field to physical atoms. In physics, energy is assigned to photons carriers 
of information, an atom by absorbing a photon receives its energy. In our social model, an excitation of the information field
just carries the social energy label $E_{\rm{c}}.$ A social atom absorbs this label and generate the corresponding portion of 
energy by itself, by transforming its psychical energy into the social energy. And the former is generated by neurophysiological 
activity in the brain and the nervous system from the physical metabolic energy. 
Thus by taking into account the psychic energy, we understand that even for cognitive systems the law of energy conservation   
is not violated.

 \end{document}